# NEPMaker: Active learning of neuroevolution machine learning potential for large cells

Junjie Wang(王俊杰),[1] Shuning Pan(潘书宁),[1] Haoting Zhang(张皓庭),[1] Qiuhan Jia(贾秋涵),[1] Chi Ding(丁驰),[1] Zheyong Fan(樊哲勇),[2] and Jian Sun(孙建)[1,*]

[1] National Laboratory of Solid State Microstructures, School of Physics and Collaborative Innovation Center of Advanced Microstructures, Nanjing University, Nanjing 210093, China

[2] College of Physical Science and Technology, Bohai University, Jinzhou, P. R. China.

## Abstract

Machine learning potentials (MLPs) achieve near first-principles accuracy but often fail for atomic environments outside the training distribution. Active learning can mitigate this limitation; however, its application to large-scale simulations is hindered by the prohibitive cost of labeling entire configurations. Here, we develop a D-optimality–driven active learning framework for the neuroevolution potential (NEP) implemented within the GPUMD package, named NEPMaker. Extrapolative atomic environments are identified on-the-fly and embedded into locally periodic structures, where boundary atoms are optimized to remain close to the training distribution. This strategy enables large-scale simulations to directly contribute to dataset construction, significantly reducing extrapolation errors while improving model robustness and transferability. The proposed framework provides a scalable route for constructing reliable machine learning potentials in complex materials systems, including those involving defects, interfaces, and phase transitions.

* Corresponding author. jiansun@nju.edu.cn
These authors contributed equally: Junjie Wang, Shuning Pan

# Introduction

Molecular dynamics (MD) simulations are a cornerstone of materials modeling, providing atomistic insight into the structural, thermal, and mechanical properties. Traditional MD typically relies on classical interatomic potentials, which are computationally efficient but often lack sufficient accuracy for complex materials. In recent years, machine learning potentials[1] (MLP) have emerged as a transformative approach. By extracting local descriptors for each atom, MLP can predict per-atom energies using linear regression[2–4], Gaussian process regression[5,6], or, most commonly, neural networks[7–12]. This combination of flexible descriptors and powerful regression models allows MLP to achieve near first-principles accuracy while maintaining computational efficiency suitable for large-scale simulations. As a result, MLP have been widely applied in MD simulations with considerable success.

Despite their high accuracy, neural network–based models remain fundamentally black boxes, and their predictions can be unreliable for atomic environments outside the training distribution[13–15]. In such cases, the model may produce significantly erroneous results, potentially leading to incorrect dynamical behavior or even simulation instability, especially in systems involving defects, phase transitions, or far-from-equilibrium processes. Consequently, reliable uncertainty quantification (UQ) is essential to guide model development and assess MLP applicability in unexplored configuration spaces[16]. Several approaches exist for estimating uncertainty. Ensemble learning[17,18] trains multiple models with different initializations, and the spread of predictions reflects the uncertainty. However, this method is relatively inefficient due to training and inferring with multiple models, and may underestimate true uncertainty. Bayesian methods[14–16], such as Gaussian process regression (GPR), provide direct uncertainty estimates, but their training complexity scales cubically with dataset size, and prediction scales quadratically. Another approach is to use distances in descriptor or latent space[19]; however, these approaches typically require additional calibration to relate distances to uncertainty, and their performance can be sensitive to the choice and scaling of descriptors, introducing further complexity.

An alternative UQ strategy for MLP is based on the D-optimality criterion[20], originally developed in the context of optimal experimental design. D-optimality quantifies the information content of a candidate training set via the determinant of the design matrix. The derived extrapolation grade directly indicates whether a given atomic environment is an interpolation of the training data or an extrapolation. D-optimality requires only a single model evaluation, enabling efficient on-the-fly identification of informative and extrapolative configurations during MD simulations. This approach has been successfully applied to MLP such as moment tensor potential (MTP)[21] and atomic cluster expansion (ACE)[22], demonstrating its effectiveness for active learning in atomistic simulations.

In this work, we implement a D-optimality-based active learning framework NEPMaker within the neuroevolution potential[23,24] (NEP) and integrate it into GPUMD[10,25] package, a GPU-accelerated MD engine optimized for large-scale simulations. The central innovation of this work is a scalable strategy for active learning in large simulation cells. Instead of labeling entire configurations, we identify and extract high-uncertainty local atomic environments from large-scale simulations on-the-fly. This allows large-scale simulations to be incorporated into the active learning process. Furthermore, in contrast to extracting atomic environments as non-periodic clusters as in MLIP-3[26], our method constructs periodic cells and optimizes the positions of atoms near the boundaries. This preserves the target atomic environment while ensuring that surrounding environments to be physically meaningful and benefit first-principles convergence. Across multiple representative systems, we demonstrate that this approach enables users to construct reliable potentials from scratch. By iteratively refining extrapolative atomic environments, the method systematically improves model stability and accuracy, ultimately enabling reliable large-scale MD simulations.

**Methods**

## A. Neuroevolution potential

In NEP potential, the local environment of atom $i$ is embedded in a descriptor vector $\{q_n^i, q_{nl}^i\}$, where $q_n^i$ is the radial part and $q_{nl}^i$ is the angular part. The descriptors depend on the chemical species and relative positions of neighboring atoms within a cutoff radius. And they are constructed to be invariant with respect to permutation, translation, and rotation. The descriptor vector is used as an input to a single layer neural network to fit the potential energy $E_i$ of the $i$-th atom:

$$E_i = \sum_{\mu=1}^{N_{\text{neu}}} w_\mu \tanh\left(\sum_{v=1}^{N_{\text{des}}} w_{\mu v} q_v^i - b_\mu\right) - b \qquad (1)$$

where $w$ and $b$ are the trainable parameters of the neural network.

## B. Estimation of uncertainty

To enable active learning, it is essential to quantify the reliability of model predictions for unseen atomic environments. In this work, we adopt a D-optimality-based criterion to estimate the extrapolation degree of atomic configurations within the NEP framework. The key idea of D-optimality is to measure how well a given atomic environment is represented by the existing training set in the descriptor space. Intuitively, environments that lie outside the span of the training data correspond to extrapolation regions, where the model predictions are less reliable.

### 1. Linear extrapolation grades

For a simple case of linear potential, the potential energy is expressed as the linear expansion of the descriptor vector:

$$E = \sum_{v=1}^{N_{\text{des}}} w_v q_v \qquad (2)$$

Suppose there are $N$ descriptors in the training set, we can put them into a matrix:

$$Q = \begin{bmatrix} q_1^1 & \cdots & q_{N_{des}}^1 \\ \vdots & \ddots & \vdots \\ q_1^N & \cdots & q_{N_{des}}^N \end{bmatrix} = \begin{bmatrix} \boldsymbol{q}^1 \\ \vdots \\ \boldsymbol{q}^N \end{bmatrix} \qquad (3)$$

Then the predicted energy $E_{predict}$ of each atom in the training set is:

$$E_{predict} = \begin{bmatrix} E_{predict}^1 \\ \vdots \\ E_{predict}^N \end{bmatrix} = \begin{bmatrix} q_1^1 & \cdots & q_{N_{des}}^1 \\ \vdots & \ddots & \vdots \\ q_1^N & \cdots & q_{N_{des}}^N \end{bmatrix} \begin{bmatrix} w_1 \\ \vdots \\ w_{N_{des}} \end{bmatrix} = QW \qquad (4)$$

The training process is equivalent to finding the weight parameters $W$ that minimize the difference between $E_{predict}$ and $E_{real}$. Since $N$ is normally much larger than $N_{des}$, we can find a subset of $Q$ that consists of $N_{des}$ descriptors to determine $W$. This subset of descriptors is called active set $A$:

$$A = \begin{bmatrix} q_1^1 & \cdots & q_{N_{des}}^1 \\ \vdots & \ddots & \vdots \\ q_1^{N_{des}} & \cdots & q_{N_{des}}^{N_{des}} \end{bmatrix} = \begin{bmatrix} \boldsymbol{q}^1 \\ \vdots \\ \boldsymbol{q}^{N_{des}} \end{bmatrix} \qquad (5)$$

The descriptors in this active set are most linear independent, and they span the largest volume in the descriptor space (have the largest determinant). Then any descriptor vectors can be expressed as a linear combination of the descriptor vector in the active set:

$$\boldsymbol{q} = \sum_{k=1}^{N_{des}} \gamma_k \boldsymbol{q}_k = \boldsymbol{\gamma} A \qquad (6)$$

Since the descriptors in the active set have the largest determinant, the absolute value of the expansion coefficients $\boldsymbol{\gamma}$ of all the descriptors in $Q$ are less or equal to one $|\gamma_k| \leq 1$. We can compute the $\boldsymbol{\gamma}$ vector of each atom by $\boldsymbol{\gamma} = \boldsymbol{q} A^{-1}$, and the active set inverse $A^{-1}$ instead of $A$ is saved during the computation process.

**2. Nonlinear extrapolation grades**

In most cases, the potential is nonlinear and the D-optimality criterion has been extended to nonlinear potentials[22,27]. The descriptor vector $q$ is replaced by another vector:

$$B = \left(\frac{\partial E}{\partial c_1}, \frac{\partial E}{\partial c_2}, \dots, \frac{\partial E}{\partial c_m}\right) \quad (7)$$

where $c$ are the trainable parameters, and $m$ is the number of all trainable parameters. The $q$ in linear potential is a special case of more general $B$ vector. For computational efficiency, we limit the trainable parameters to neuron network parameters, and neglect the parameters of the radial functions in the descriptor[22]. Specifically, according to Eq. 1 the components of $B$ for an NEP potential is:

$$B_{w_\mu} = \frac{\partial E}{\partial w_\mu} = \tanh\left(\sum_{v=1}^{N_{des}} w_{\mu v} q_v - b_\mu\right) \quad (8)$$

$$B_{w_{\mu v}} = \frac{\partial E}{\partial w_{\mu v}} = w_\mu q_v \left(1 - \tanh^2\left(\sum_{v=1}^{N_{des}} w_{\mu v} q_v - b_\mu\right)\right) \quad (9)$$

$$B_{b_\mu} = \frac{\partial E}{\partial b_\mu} = -w_\mu \left(1 - \tanh^2\left(\sum_{v=1}^{N_{des}} w_{\mu v} q_v - b_\mu\right)\right) \quad (10)$$

After obtaining the $B$ vector of all the atoms in the training set, an active set is selected with the MaxVol algorithm[28]. We implement the MaxVol algorithm with the CuPy package[29], which can be executed with high efficiency on GPU.

### 3. Performance of D-optimality criterion

To validate the effectiveness of the D-optimality criterion within the NEP framework, we applied it to the GAP dataset of silicon[30]. The GAP dataset consists of various silicon phases, including diamond phase, $\beta$-Sn phase, hexagonal phase, etc. The NEP model is trained using 90% of the configurations from the diamond and β-Sn phases, while the remaining configurations are used for testing. Additional silicon phases that are not included in the training set are used to assess extrapolation behavior.

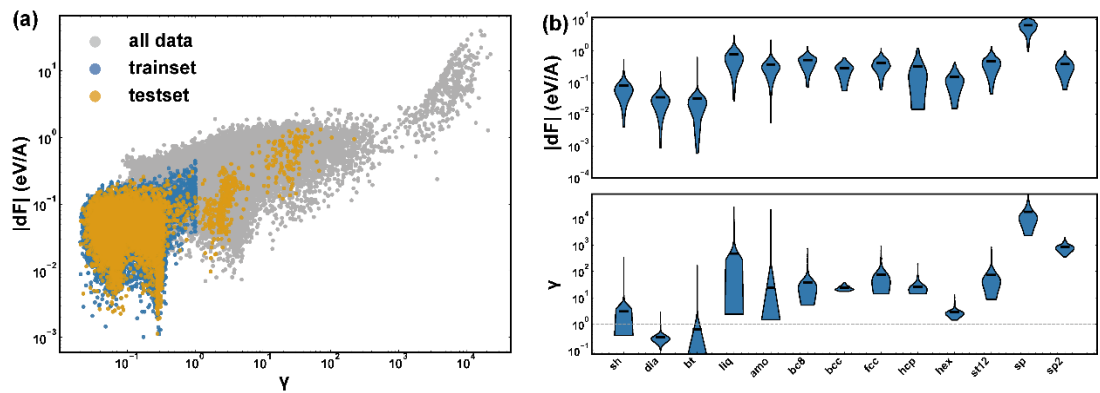

Fig 1. (a) Force errors versus the extrapolation grade in D-optimality for the Si datasets. Blue and yellow points represent the training and test sets respectively, containing configurations from the diamond and β-Sn phases only. Gray points denote the full dataset, including all phases. (b) Violin plots of force errors and extrapolation grades across different types of structures.

Fig. 1(a) shows the relationship between the force prediction error and the extrapolation grade $\gamma$. A clear positive correlation between $\gamma$ and the prediction error is observed. For configurations in the training set, all $\gamma$ values are smaller than 1, indicating that these environments are well represented by the training data. In contrast, a subset of configurations in the test set exhibits larger $\gamma$ values accompanied by increased prediction errors, suggesting that these environments lie outside the training domain. For the silicon phases that are entirely absent from the training set marked as grey points in Fig .1(a), both $\gamma$ values and prediction errors are significantly larger. The distributions of $\gamma$ and force errors for different phases are further illustrated in Fig. 1(b). The results consistently show that structures with larger $\gamma$ values tend to have higher prediction errors across all phases. These results demonstrate that the extrapolation grade $\gamma$ provides a reliable and physically meaningful indicator of model uncertainty, justifying its use in the active learning framework.

**4. Multiple elements**

For multi-element materials, the NEP potential employs different parameters for different elements, and the atomic environments of different elements naturally differ. As a result, the values of the **B** vectors can vary by several orders of magnitude. Directly applying the MaxVol algorithm to the combined **B** matrix of all atoms may therefore lead to biased selection, where atomic environments from certain elements dominate while others are underrepresented. A possible solution is to first cluster the **B** vectors and then apply the MaxVol algorithm within each cluster[31,32]. However, in the context of large-scale active learning, such clustering requires storing all candidate environments and introduces substantial computational and implementation overhead. Instead, we adopt a simpler and more efficient strategy by partitioning atomic environments according to their chemical species and applying the MaxVol algorithm independently to each group. This approach effectively alleviates the scale imbalance in the B vectors while ensuring a balanced selection across different elements.

## C. Local fragment extraction

To combine active learning with large-scale MD simulations, it is essential to generate physically meaningful first-principles data for extrapolative atomic environments identified in large simulation cells. In MLIP-3[26], the target atomic environment is extracted based on a cutoff radius and placed in vacuum. However, this strategy leads to non-physical structures and introduces ambiguity in atomic energy assignment due to contributions from out-of-distribution environments. An alternative approach proposed by Jalolov *et al.*[33] constructs periodic configurations from cut fragments, followed by first-principles relaxation of atoms outside the extrapolation region while keeping the target environment fixed. This ensures that the extrapolative atomic environment remains unchanged during relaxation. This approach improves the physical consistency of the generated structures compared to vacuum-based cluster methods, and can enhance the stability and accuracy of MLP training. However, it requires additional first-principles relaxation, leading to increased computational cost. To address this problem, In-Distribution substructure Embedding Active Learner

(IDEAL)[34] optimizes the surrounding environment using an uncertainty-based criterion rather than first-principles energies. This strategy reduces the reliance on expensive first-principles calculations while constraining the surrounding environments to remain close to the training distribution, thereby reducing ambiguity in atomic energy assignment and improving both the stability and accuracy of model training.

Here, we adopt a strategy similar to IDEAL. We first extract the region containing the target atomic environment from the whole structure and then optimize the atoms outside this region by minimizing an uncertainty-based objective. The main difference is that IDEAL first determines the lattice via random search and keeps it fixed during subsequent optimization, whereas in our approach, the lattice and atomic positions are optimized simultaneously. Moreover, IDEAL keeps the atoms within the local atomic environment fixed, but during optimization, external atoms may enter this cutoff region, causing the atomic environment to change. We avoid this by introducing an additional hard-sphere potential. Additionally, IDEAL employs M3GNet[35] for the MD simulations and estimates uncertainty using SOAP descriptors calculated by DScribe package[36] combined with the Mahalanobis distance. In contrast, our method computes uncertainty consistently with the MLP used in the simulation, using the $\gamma$ vector obtained from NEP for uncertainty estimation as:

$$\text{Unc}_p = \sum_i \max(\|\gamma_i\|_p, 1) \tag{11}$$

where $\|\gamma\|_p = \left(\sum_d \gamma_d^p\right)^{\frac{1}{p}}$ is p-norm of $\gamma$ vector to obtain a smooth approximation of $\max(\gamma)$, and we set p=6 in our subsequent calculations. The use of $\max(\|\gamma_i\|_p, 1)$ ensures that atomic environments within the interpolation regime do not contribute to further optimization. Without this constraint, the optimization would tend to artificially drive $\gamma$ values toward zero, leading to non-physical configurations.

To demonstrate the optimization of boundaries and its importance in our method,

we consider the silicon system described above. A NEP potential with a cutoff radius of 3 Å is trained using all non-liquid structures and applied to a large-cell liquid silicon dataset. For the extrapolated liquid atomic environments identified using gamma as a criterion, we first extracted a cubic region of 5 Å side length centered on each environment and added a vacuum layer. As shown in Fig. 2b, this naive truncation leads to artificially large γ values for boundary atoms, deviating significantly from those in the original liquid environment. This effect is further illustrated in the principal component analysis (PCA) map of the local atomic descriptors (Fig. 2c), where truncated structures produce outlier environments far from the training distribution. After applying our optimization procedure, the target local environments remain unchanged as indicated by the stable gamma values of the central atoms, while the uncertainties of boundary atoms are reduced (Fig. 2d). This results in physically more consistent structures and avoids the introduction of out-of-distribution environments into the active learning dataset.

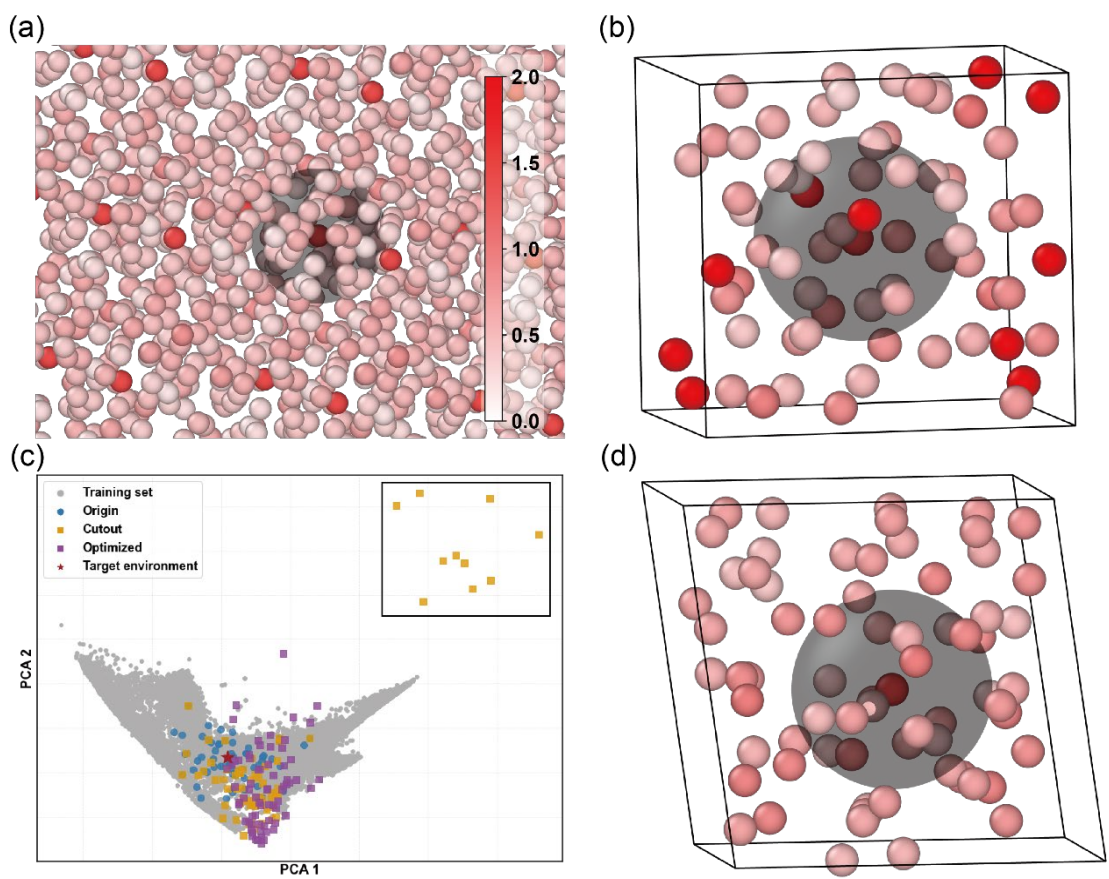

Fig 2. Uncertainty-based optimization of boundary atomic environments. Extrapolation grade is computed using an NEP potential trained on all non-liquid structures, with atoms colored from white to red according to their grade. (a) Target environment identified in a liquid supercell. (b) A cutout cell with added vacuum, where periodic boundaries introduce spurious extrapolative environments. (c) PCA visualization of atomic descriptors: gray points (training set), blue circles (liquid), red stars (identified extrapolative points), yellow squares (raw cutout), and purple squares (after uncertainty-based optimization). Outliers appear in the upper-right region before optimization. (d) Optimized structure preserving environments within the cutoff while adjusting atoms outside the sphere and the lattice to reduce overall uncertainty.

### D. Framework of NEP active learning

The analysis above demonstrates that the D-optimality criterion provides a reliable quantitative indicator for identifying extrapolative atomic environments within the NEP framework. Moreover, for locally periodic fragments extracted from large supercells, uncertainty-driven structural optimization preserves the target local environment while guiding surrounding atoms toward configurations consistent with the training distribution. Based on these observations, we construct an automated active-learning workflow for NEP potentials as shown in Fig.3, consisting of the following steps:

(1) **Initial training.** An initial training set is prepared, either manually or from prior datasets, to fit a baseline NEP potential. Subsequent iterations update the training set automatically.

(2) **Active set construction.** The MaxVol algorithm is applied to the current training dataset to select a representative active set, which serves as the reference for extrapolation detection.

(3) **Exploration.** MD simulations are performed under target thermodynamic conditions, and configurations with large $\gamma$ values are identified as candidates for extrapolation.

(4) **Selection and refinement.** Extrapolative configurations are merged with the existing dataset, followed by a second MaxVol selection to identify informative structures. A random down-selection can be applied to control the number of added samples. For large supercell simulations, local atomic environments are further extracted and their boundaries optimized to ensure physical consistency.

(5) **First-principles labeling.** Density functional theory (DFT) calculations are carried out for the selected structures, and the resulting data are incorporated into the training set.

Steps (1) – (5) are repeated until no extrapolative environments are detected.

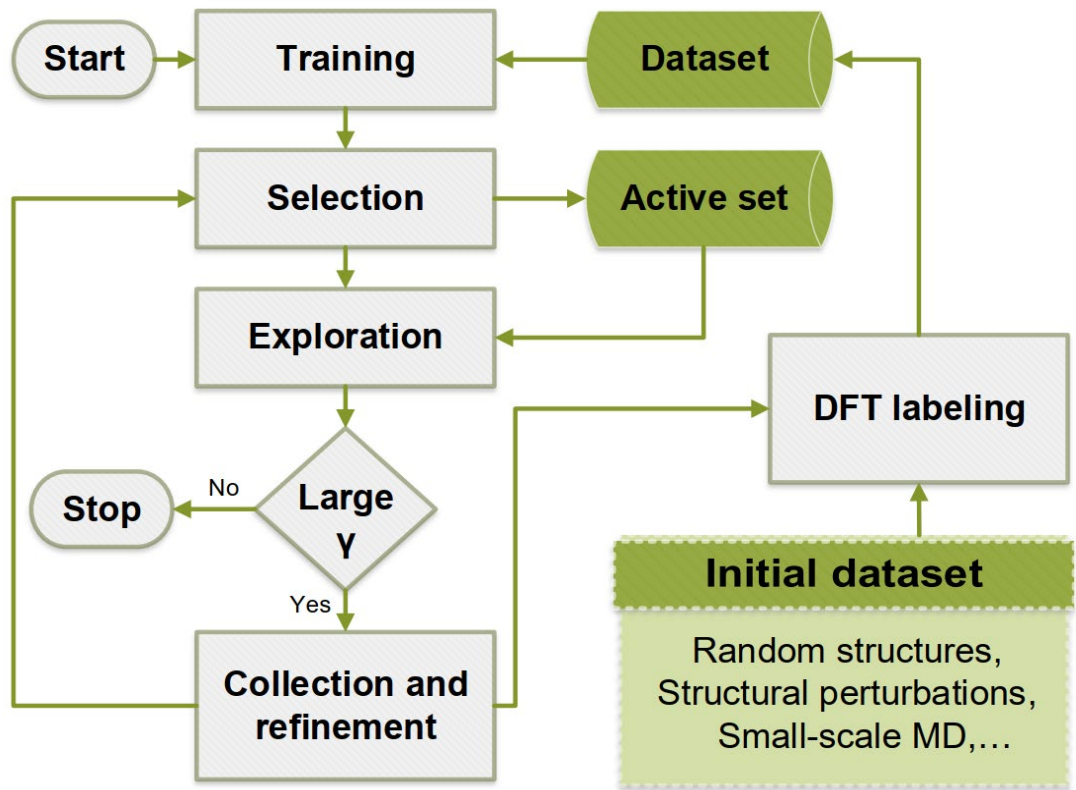

Fig 3. Framework of NEP active learning. Starting from an initial training set, a NEP potential is iteratively improved through exploration, selection, and retraining. Extrapolative atomic environments are identified using the γ-based uncertainty criterion during MD simulations. For large-scale systems, local environments are extracted and optimized to ensure physical consistency. Selected structures are labeled by first-principles calculations and incorporated into the training set until convergence.

**Results**

**A. Melting of sodium**

We first validate the active learning procedure on the melting of sodium in small cell. Sodium is a simple system suitable for demonstrating the workflow. As a simple metallic system with a single valence electron, sodium enables computationally efficient first-principles calculations while retaining essential physics. The initial training set consists of 20 slightly perturbed solid Na structures. MD simulations were performed on a 250-atom system, which was first heated from 250 K to 500 K and then cooled back to 250 K. Since the system size is within the feasible range for DFT, full first-principles calculations were performed on the selected structures without local cutouts. To accelerate sampling, eight MD trajectories with different random seeds were run in parallel.

In each active learning iteration, up to 20 new structures were added to the training set. As shown in Fig. 4a, the number of completed MD steps increases with iteration number as the training set grows. By the sixth iteration, all MD trajectories reach completion, although some extrapolative structures with high $\gamma$ values still appear (Fig. 4b) and should be added to the training set. Notably, the number of extrapolative structures added in Fig. 4b exhibits a non-monotonic trend, first increasing and then decreasing. This behavior arises because, in the early stages of active learning, simulations terminate rapidly due to model instability, resulting in limited sampling. As the model improves, longer trajectories enable more extensive exploration and thus the identification of a larger number of extrapolative configurations. In the later iterations, the decline reflects the progressive convergence of the potential, with fewer configurations exceeding the extrapolation threshold. In the seventh iteration, all MD trajectories run to completion, and no structures exceed the predefined extrapolation threshold $\gamma_{low}$. The active learning loop converges after seven iterations, yielding a final training set of 130 structures. The NEP potential obtained at each iteration was evaluated against the final training set, and the root-mean-square errors (RMSE) were found to decrease and converge with increasing iteration number (Fig. 4c).

The converged NEP potential was further validated by calculating the melting point of sodium at 0 GPa using the two-phase method. Coexistence cells containing 20,000 atoms with both solid and liquid regions were constructed for temperatures ranging from 330 K to 380 K. For each temperature, the system was equilibrated at the target pressure, the solid region was melted at 500 K, and the liquid was cooled back to the target temperature, followed by a 1 ns NPH simulation to allow the solid and liquid phases to reach equilibrium. The temperature at which the solid and liquid coexist stably was identified as the melting point. The calculated melting point under ambient pressure is around 350 K, which is closed to the experimental value 370 K.

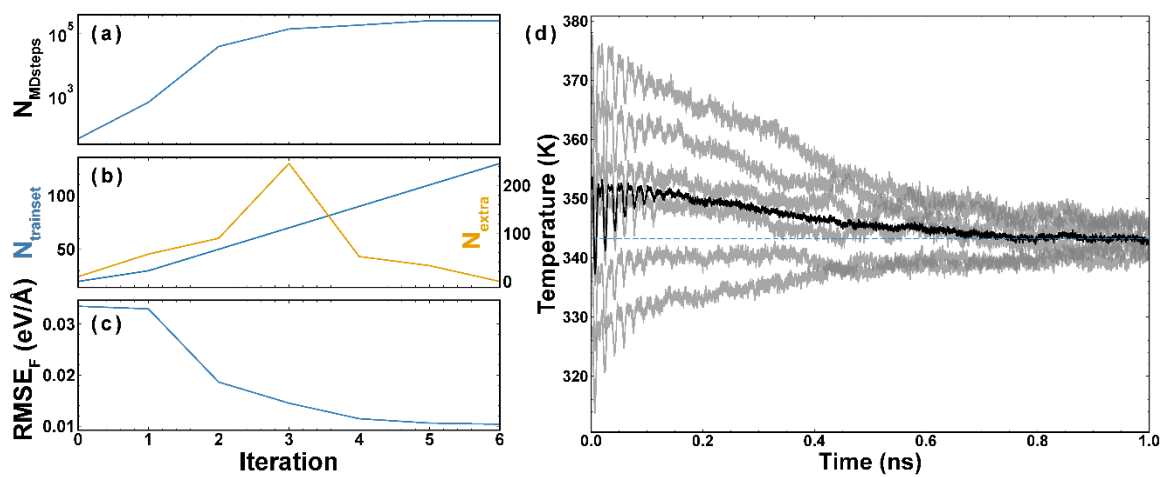

Fig 4. Analysis of the active learning procedure and melting-point calculations for sodium. (a–c) Quantities shown as a function of iteration: (a) Number of successfully completed MD steps. (b) Evolution of the training-set size and number of structures exceeding the extrapolation threshold $\gamma$. (c) Force RMSE of the NEP model on the final training set. (d) Temperature evolution in NPH simulations initiated at different temperatures.

**B. Solid-solid phase transitions in CsPbI₃**

We then apply the active learning procedure to the inorganic halide perovskite CsPbI₃, a prototypical system exhibiting pronounced anharmonic lattice dynamics and multiple temperature-driven structural phase transitions. Similar to other halide perovskites, the perovskite phases of CsPbI₃ are connected through soft phonon modes, making static calculations insufficient to describe its finite-temperature behavior. Accurate characterization of the phase transitions therefore requires MD simulations capable of accessing long time scales and sufficiently large system sizes. CsPbI₃ crystallizes in three perovskite phases: the high-temperature cubic α phase, the intermediate tetragonal β phase, and the low-temperature orthorhombic γ phase, which transform into one another upon heating and cooling[37]. We therefore take the low-temperature γ phase as the initial structure, heat the system from 20 K to 620 K, and subsequently cool it back to 20 K. The active learning procedure converges after 23 iterations, yielding a training set comprising 401 structures.

With the converged potential, we perform MD simulations with larger cells to check the phase transition temperature. A supercell containing 23 040 atoms was constructed by replicating the orthorhombic γ-phase unit cell, which has been shown in previous studies[38,39] to effectively mitigate size effects and yield well-converged results, enabling the explicit treatment of lattice distortions and anisotropic strain associated with the phase transitions. The system was first equilibrated at 100 K under the NPT ensemble and then continuously heated from 100 K to 600 K in 5ns with fully flexible cell degrees of freedom, allowing the lattice parameters to evolve naturally with temperature. From the evolution of the lattice constants according to temperature as shown in Fig. 5, we observe that the dynamical trajectories generated by the present potential correctly reproduce the solid–solid phase transition sequence. With increasing temperature, the lattice constants *a* and *c* increase while *b* decreases. Around 280 K, where *a* = *b*, the system undergoes a transition from the orthorhombic *γ* phase to the tetragonal *β* phase. Upon further heating, *a* and *b* continue to increase whereas *c* decreases, until all three lattice constants converge to the same value at approximately 400 K, indicating a transition into the cubic $\alpha$ phase. These results are in good agreement with previously reported transition temperatures and pathways, demonstrating that a physically meaningful and transferable interatomic potential has been successfully learned from scratch.

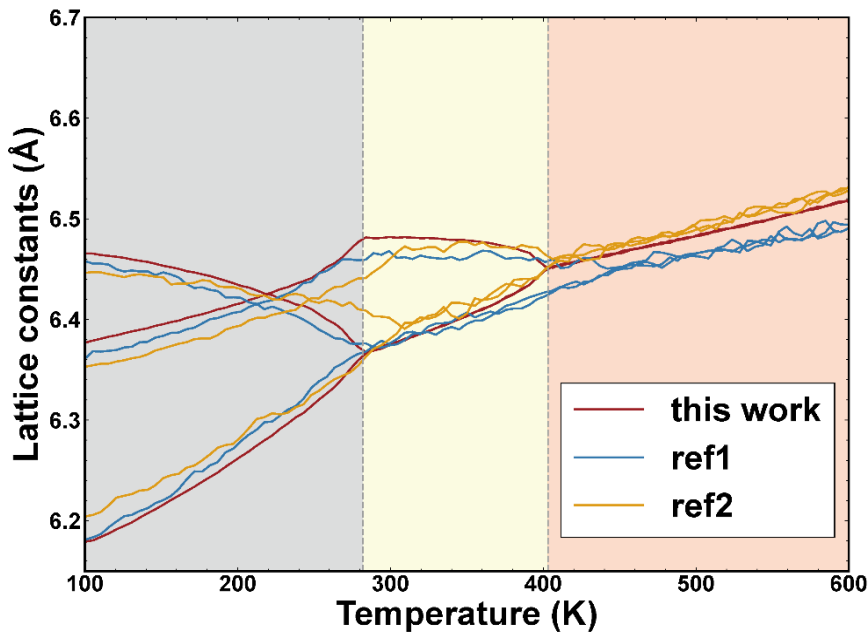

Fig 5 Lattice parameters as a function of temperature from simulation. The red, blue, and yellow lines correspond to the simulation results obtained by this work, Fransson (ref 1)[38], and Chen (ref 2)[39], respectively. All these data were generated using NEP models in combination with GPUMD simulations and labeled with PBE calculations.

### C. B4-B1 phase transitions in GaN

Finally, we employ the large-supercell active learning approach to investigate the solid-solid phase transition in GaN. This system exhibits pronounced size effects during the B4–B1 phase transition[40–42], making it an ideal benchmark for assessing the capability of large-scale simulations and the effectiveness of active learning strategies. The proposed active learning framework enables on-the-fly sampling and model updates directly during MD simulations of large supercells. This allows the active learning procedure to naturally capture complex configurations emerging during the phase transition, including local lattice distortions, phase interfaces, and inhomogeneous strain, thereby alleviating the biases associated with conventional small-scale sampling.

The initial dataset consisted of 200 random GaN configurations[41]. Metadynamics simulations were conducted at 300 K and 50 GPa using GPU-MetaD[43], starting from a B4-phase GaN supercell containing 2048 atoms, with coordination number and volume selected as collective variables (CVs) to facilitate the B4-B1 phase transition. Compared to the previous simulation starting from scratch, the number of iterations was reduced by employing random structures as initial configurations. The active learning procedure converged after 12 iterations, yielding a training set of 1565 structures. Starting from this model, we subsequently apply the workflow to a larger supercell containing 27,648 atoms. Due to size effects, distinct transition pathways emerge in the larger system, leading to additional extrapolative configurations during metadynamics. The active learning process converges after 9 iterations, resulting in a final training set of approximately 2,000 structures.

Since the simulations were performed directly with large supercells, the target MD trajectories are obtained simultaneously with model construction. In the simulation with 2048 atoms, we identify two distinct transformation mechanisms from the metadynamics trajectories as shown in Fig.6. In the first pathway (a-$b_1$-$c_1$-d), the hexagonal rings undergo reconstruction within the plane normal to the $b$ axis, which initiates the formation of a B4-B1 interface. This interface subsequently propagates through the lattice, ultimately transforming the entire supercell into the B1 phase. Alternatively, the second pathway (a-$b_2$-$c_2$-d) proceeds via compression along the c axis, generating a metastable intermediate structure characterized by five-fold-coordinated atoms. Bond breaking and reformation then occur within the plane normal to the c axis, driving the system toward the final B1 configuration. In the 27,648-atom supercell simulations, the crystal lattice experiences cyclic elongations and contractions, eventually yielding a reticulated pattern of striated domains populated by 5-fold-coordinated atoms, which is consistent with prior studies conducted under analogous thermodynamic conditions[40]. Subsequently, nucleation sites featuring 6-fold-coordinated atomic environments develop within this mesh-like architecture. These results illustrate our ability to execute active learning procedures in large supercells intractable for first-principles simulation, extracting extrapolative atomic environments on-the-fly and bringing simulations to completion within these extended systems.

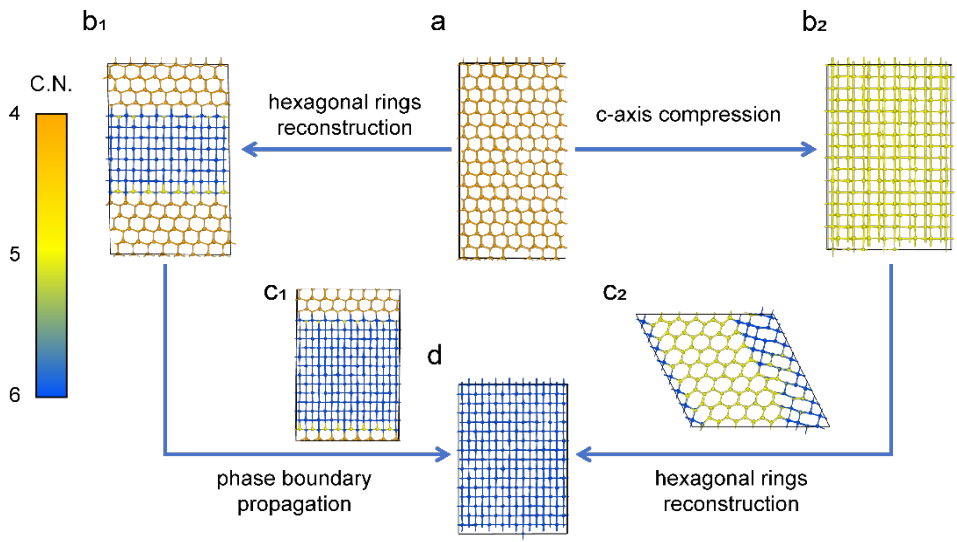

Fig 6. The pathways of GaN B4-B1 phase transitions with simulation cells containing 2 thousand atoms. (a-b1-c1-d) path shows hexagonal rings reconstruction in plane normal to b axis, forming a B4-B1 interface, which then propatgation and turning the cell into B1 phase. In (a-b2-c2-d) path the cell goes through a c-axis compression, forming a 5-fold-coordinated intermediate structure. Bond breaking and reformation then occur in plane normal to c axis, resulting in the final transformation to the B1 phase.

**Computational Details**

All first-principles calculations were performed using the Vienna ab initio simulation package (VASP)[44]. The projector augmented-wave (PAW) together with the Perdew–Burke–Ernzerhof (PBE)[45,46] exchange correlation functional was employed.

Brillouin zone sampling was carried out using a Γ-centered k-point grid with a spacing less than 0.5 Å$^{-1}$. A plane-wave energy cutoff of 125 eV, 220 eV and 520 eV was used for Na, CsPbI$_3$, and GaN.

To train the machine learning potentials, the NEP4 version was employed. For all systems, the basis sizes for radial and angular functions were both set to 8. For Na and CsPbI$_3$, the radial and angular cutoffs were set to 8 Å and 4 Å, respectively. The maximum angular momentum quantum numbers were chosen as l$_{max}$ = 4, 2, and 0 for the three-body, four-body, and higher-order terms, respectively. For GaN, the radial and angular cutoffs were set to 4 Å and 3 Å, and the corresponding maximum angular momentum quantum numbers were l$_{max}$ = 4, 0, and 0. Upon convergence of the active learning iterations, the resulting NEP models achieve RMSE for energy per atom, forces, and virials of 0.41 meV/atom, 10.38 meV/Å, and 3.78 meV/atom for Na; 0.69 meV/atom, 43.38 meV/Å, and 6.67 meV/atom for CsPbI$_3$; and 15.24 meV/atom, 274.44 meV/Å, and 87.15 meV/atom for GaN, respectively.

**Conclusions**

In this work, we propose and validate a D-optimality-based active learning framework that enables the training dataset of the neuroevolution potential (NEP) to be continuously and automatically refined during structural exploration with GPUMD. By identifying and selectively extracting atomic environments with high uncertainty on the fly, this approach achieves systematic and efficient iterative optimization of machine-learned interatomic potentials without the prohibitive cost of retraining on entire large-scale configurations. The introduction of a periodic atomic environment extraction strategy with optimized boundary atoms further improves the convergence of first-principles calculations and effectively minimizes the interference of surrounding environments on the target atomic environments. Moreover, our workflow can be seamlessly integrated into ongoing MD simulations, enabling automated, on-the-fly potential refinement. The robustness and transferability of the proposed workflow are demonstrated across three representative systems, including the melting of sodium, solid-solid phase transitions in CsPbI$_3$, and the B4–B1 phase transition in GaN. In all cases, the resulting NEP models reproduce the reported phase transition pathways and characteristic temperatures, confirming the reliability of the active learning strategy. Overall, this framework provides a general and scalable solution for the automated, on-the-fly construction of machine learning potentials in complex dynamical processes, combining first-principles accuracy with uncertainty control in large-scale MD simulations.

## Acknowledgements

This work was supported by the National Natural Science Foundation of China (Grants No.12125404, T2495231, and 12504277), the National Key R&D Program of China (Grant No. 2022YFA1403201), the Advanced Materials-National Science and Technology Major Project (Grant 2024ZD0607000), the Basic Research Program of Jiangsu (Grants BK20233001, BK20241253), the Jiangsu Funding Program for Excellent Postdoctoral Talent (Grants 2024ZB002, 2024ZB075), the Postdoctoral Fellowship Program of CPSF (Grant GZC20240695), Fundamental and Interdisciplinary Disciplines Breakthrough Plan of the Ministry of Education of China (JYB2025XDXM413), Science Challenge Project (No. TZ2025013), the AI & AI for Science program of Nanjing University, Artificial Intelligence and Quantum physics (AIQ) program of Nanjing University, and the Fundamental Research Funds for the Central Universities. The calculations were carried out using supercomputers at the High-Performance Computing Center of Collaborative Innovation Center of Advanced Microstructures, the high-performance supercomputing center of Nanjing University.

## Code Available

The NepMaker package is freely available at https://gitlab.com/bigd4/activemiao.